# Grand Challenges for Global Brain Sciences

Global Brain Workshop 2016 Attendees*

The next grand challenges for science and society are in the brain sciences. A collection of 60+ scientists from around the world, together with 15+ observers from national, private, and foundations, spent two days together discussing the top challenges that we could solve as a global community in the next decade. We settled on three challenges, spanning anatomy, physiology, and medicine. Addressing all three challenges requires novel computational infrastructure. The group proposed the creation of The International Brain Station (TIBS), to address these challenges, and launch brain sciences to the next level of understanding.

Understanding the brain and curing its diseases are among the most exciting challenges of our time. Consequently, national, transnational, and private parties are investing billions of dollars (USD). To efficiently join forces, *Global Brain Workshop 2016* was hosted at Johns Hopkins University's Kavli Neuroscience Discovery Institute on April 7-8. A second workshop, *Open Data Ecosystem in Neuroscience* took place July 25-26 in Washington, DC to continue the discussion specifically about computational challenges and opportunities. A third conference, *Coordinating Global Brain Projects,* took place in New York City on September 19th in association with the United Nations General Assembly. So vast are both the challenges and the opportunities that global coordination is crucial.

To find ways of synergistically studying the brain, the kick-off workshop welcomed over 60 scientists, representing 12 different countries and a wide range of subdisciplines. They were joined by 15 observers from various national and international funding organizations. Participants were engaged weeks before the conference and charged with coming up with ambitious projects that are both feasible and internationally inclusive, on par with the International Space Station (i.e., worthy of a global, decade-long effort). Over the course of 36 hours, scientists discussed, debated, and gathered feedback, ultimately proposing several "grand challenges for global brain sciences" that were refined by working groups. The workshop was covered in a media piece in *Science* April 15, 2016.

The group began with 60+ ideas, each forged independently by one of the scientific participants. Each participant proposed a unique challenge that was designed to meet the following desiderata:

1. *Significant*: it will yield tangible societal, economic, and medical benefits to the world.

2. *Feasible*: it can achieve major milestones within 10 years given existing funding opportunities.
3. *Inclusive*: nations throughout the world can meaningfully contribute to and benefit from each challenge, and the collection of challenges are collectively scientifically diverse.

Interestingly, a lot of the proposed ideas were similar to one another and others were complementary. This allowed the group to converge on three grand challenges for global brain sciences, each depending on a common universal resource.

## Challenge 1: What makes our brains unique?

Both within and across species, brain structure is known to exhibit significant variability across many orders of magnitude in scale—*including anatomy, biochemistry, connectivity, development, and gene expression* (ABCDE). It remains mysterious how and why the nervous system tightly regulates certain properties, while allowing others to vary. Understanding the design principles governing variability may hold the key to understanding intelligence and subjective experience, as well as the influence of variability on health and function.

**This grand challenge is a global project to coordinate the construction of comprehensive multiscale maps of the ABCDE's of multiple brains from multiple species using multiple cognitive and mental health disease models**. Within a decade, we expect to have addressed this challenge in brains including but not limited to Drosophila, Zebrafish, Mouse, and Marmoset, and to have developed tools to conduct massive neurocartographic analyses. The result will be a state-of-the-art "Virtual NeuroZoo" with fully annotated data and analytic tools for analysis and discovery. This virtual NeuroZoo can be utilized by neuroscientists and citizens alike, both as a reference and for educational materials. By incorporating disease models, we explicitly link this challenge with the third challenge.

## Challenge 2: How does the brain solve complex computational problems?

Brains remain the most computationally advanced machines for a large array of cognitive tasks—whether navigating hazardous terrain, translating languages, conducting surgery, or recognizing emotional states—despite the fact that modern computers can utilize millions of training samples, megawatts of power, and tons of hardware. While the ABCDEs establish the "wetware" upon which our brains can solve such computations, to understand the mechanisms we need to measure, manipulate, and model neural activity simultaneously across many spatiotemporal resolutions and scales—including wearables, embedded sensors, and actuators—while animals are exhibiting complex ecological behaviors in naturalistic environments.

**This grand challenge is a global project to investigate a single naturalistic behavior that is ecologically relevant across phylogenies, such as foraging, and measure brain and body properties across spatial, temporal, and genetic scales.** The challenge differs from previous efforts in three key ways. First, it requires studying animals in *complex and naturalistic* environments. Second, it requires *coordinated* attacks at many different scales by many different investigators while the animals are performing the same complex behaviors. We envision groups of 20-30 investigators all operating together to share data and experimental design. Third, the richness of the mental repertoire of cognition suggests that deciphering its codes will require *many parallel investigations* to uncover different facets of brain function. These experiments in turn will produce multiscale models of neural systems with the potential to accomplish computational tasks that no current computer system can perform. Mechanistic studies, guided by theoretical models, will help to ask how perturbations of those systems lead to aberrant function, linking this challenge with the next one.

## Challenge 3: How can we augment clinical decision-making to prevent disease and restore brain function?

Psychiatric and neurological illnesses levy enormous burdens upon humanity: impairment, suffering, financial costs, and loss of productivity. Despite a growing awareness of the challenges, clinicians consistently battle the lack of objective tests to guide clinical decision-making (e.g., diagnosis, selection of treatments, prognosis). Compounding these limitations are societal stigmas regarding mental illness that increase the suffering of patients and their families. The ABCDEs of neurobiological variability, when coupled with multiscale mechanistic models of cognition, will provide new approaches to neurobiologically-informed clinical decision making.

**This grand challenge is a global project to transform clinical decision-making via incorporating neural mechanisms of dysfunction.** This will require collecting, organizing and analyzing human and non-human anatomical and functional data. These data, and the tools developed to explore and discover novel treatment therapies, will be the foundation upon which the next decades of experiments and clinical decisions will be based. The distributed and multimodal nature of these datasets further motivate the need for an all-purpose computational platform, upon which models of disease can be developed, deployed, tested, and refined.

## A Universal Resource

All three of the grand challenges for global brain sciences represent severe methodological challenges, both technological and computational. The technological developments required for each of the challenges are non-overlapping. In contrast, regardless of the nature of the scientific questions or data modalities involved, each project will require computational capabilities including collecting, storing, exploring, analyzing, modeling, and discovering data. Although neuroscience has developed a large number of computational tools to deal with existing datasets, the datasets proposed here bring with them a whole suite of new challenges.

This resource would be a comprehensive computational platform, deployed in the cloud, that will provide web services for all the current “pain points” in daily neuroscience practice associated with big data. This resource will realize a new era of brain sciences, one in which the bottlenecks to discovery transition away from data collection and processing to data enriching exploring, and modeling. While science has always benefitted from standing on the shoulders of giants, this *will enable science to stand on the shoulders of everyone*. Today, essentially every practicing neuroscientist’s productivity is limited due to computational resources, access to data or algorithms, or struggling with determining which data and algorithms are best suited to answer the most pressing questions of our generation. This resource will create a future where those limitations will feel as archaic as fitting the data with paper and pencil feels today. For further details, see an upcoming NeuroView called “To the Cloud! A Grassroots Proposal to Accelerate Brain Science Discovery”.

## Societal Considerations

Each nation affords different opportunities and restrictions, owing to ethical, policy, and cultural considerations. Because these grand challenges are inherently inclusive, manifesting them will require understanding and mitigating issues that arise in cross-cultural endeavors. Indeed, addressing the vast diversity of partnerships in such an endeavor is a challenge in itself. We therefore recommend the following. First, form a **cultural sensitivity committee** to consider and investigate potentially sensitive issues. Second, bolstered by their research, establish **cross-cultural collaboration education materials**, including written guidelines and videos, which will be recommended to all participating scientists. Third, to deepen the understanding of transnational collaborations, develop **trainee exchange programs** in which participating trainees will spend six months to a year working and training in a foreign country. This will also facilitate cross-cultural knowledge dissemination and fertilization. Fourth, require **frequent assessments** to ensure maintenance of cultural sensitivities. These assessments will feedback into the educational material and be used to modify the exchange programs.

## Next Steps

Crucial to the success of this endeavor is a sequence of actionable steps that the community can follow. Because we are not proposing any additional funding, realizing the eventual goals of these grand challenges will rely on marshalling existing funds. Due to the incoming leadership changes, both on national and transnational levels, quick action is of the essence. Therefore, we have taken the following steps: We have created a webpage, http://brainx.io, containing a bibliography of reports that resulted from this conference, as well as a list of all scientific participants and observers who attended the original brainstorming meeting leading to this document. We will also be monitoring comments on https://neurostars.org/ with the tag “neurostorm” for further discussion. Finally, we will have an outpost at the NeuroData booth (#4126) at the SfN meeting in San Diego to discuss these issues further. We encourage anybody who feels inspired by this document to join the discussion, engage, and get in touch with funders and other scientists with your ideas.

**Acknowledgements**

National Science Foundation (1637376) and the Kavli Foundation.

***Global Brain Workshop 2016 Attendees**

- Joshua T. Vogelstein[1,27,28,29,30,31]
- Katrin Amunts[7,8]
- Andreas Andreou[30]
- Dora Angelaki[32]
- Giorgio A. Ascoli[33]
- Cori Bargmann[34]
- Randal Burns[28, 29]
- Corrado Cali[11]
- Frances Chance[35]
- George Church[36]
- Hollis Cline[37]
- Todd Coleman[38]
- Stephanie de La Rochefoucauld[39]
- Winfried Denk[40]
- Ana Belén Elgoyhen[41]
- Ralph Etienne Cummings[42]
- Alan Evans[5]
- Kenneth Harris[43]
- Michael Hausser[3]
- Sean Hill[9]
- Samuel Inverso[44]
- Chad Jackson[45]
- Viren Jain[46]
- Rob Kass[47]
- Bobby Kasthuri[13]
- Adam Kepecs[15]
- Gregory Kiar[1, 27]
- Dean M. Kleissas[25]
- Konrad Kording[6]
- Sandhya P. Koushika[10]
- John Krakauer[48]
- Story Landis[49]
- Jeff Layton[50]
- Qingming Luo[51]
- Adam Marblestone[52]
- David Markowitz[26]
- Justin McArthur[53]
- Brett Mensh[2,4]
- Michael P. Milham[19]
- Partha Mitra[15]
- Pedja Neskovic[54]
- Miguel Nicolelis[55]
- Richard O'Brien[56]
- Aude Oliva[57]
- Gergo Orban[58]
- Hanchuan Peng[14]
- Eric Perlman[27]
- Marina Picciotto[59]
- Mu-Ming Poo[17]
- Jean-Baptiste Poline[18]
- Alexandre Pouget[60]
- Sridhar Raghavachari[61]
- Jane Roskams[14]
- Alyssa Picchini Schaffer[20]
- Terry Sejnowski[62]

- Friedrich T. Sommer[63]
- Nelson Spruston[4]
- Larry Swanson[64]
- Arthur Toga[65]
- R. Jacob Vogelstein[26]
- Anthony Zador[15]
- Richard Huganir[30,31]
- Michael I. Miller[1,27,31]

1. Department of Biomedical Engineering, Institute for Computational Medicine, Johns Hopkins University, Baltimore, MD, USA
2. Optimize Science, Mill Valley, CA USA; UCSF Kavli Institute for Fundamental Neuroscience, San Francisco, CA, USA
3. Department of Physiology, University College London, London, UK
4. Janelia Research Campus, Howard Hughes Medical Institute, Ashburn, VA, USA
5. Montreal Neurological Institute, McGill University,,Montreal, Quebec, Canada
6. Physical Medicine and Rehabilitation, Physiology, and Applied Mathematics, and Biomedical Engineering, Northwestern University, Chicago, IL, USA
7. Institute for Neuroscience and Medicine, INM-1, Research Centre Juelich, Germany, C. and O. Vogt Institute for Brain Research, Forschungszentrum Jülich; University Hospital Duesseldorf, University Duesseldorf, Germany
8. Human Brain Project, EPFL, Geneva, Switzerland
9. Blue Brain Project, EPFL, Campus Biotech, Geneva, Switzerland
10. Department of Biological Sciences, Tata Institute of Fundamental Research, Navy Nagar, Colaba, Mumbai, India
11. Biological and Environmental Science and Engineering, KAUST,Thuwal, 23955-6900 Saudi Arabia
12. Cuban Neuroscience Center, 190 e / 25 and 27, Cubanacan, Playa. Havana. CP 11600; University of Electronic Science and Technology of China, Shahe Campus:No.4, Section 2, North Jianshe Road, 610054, Chengdu, Sichuan, P.R.China
13. Argonne National Laboratory, Argonne, IL, USA
14. Allen Institute for Brain Science, Seattle, WA, USA
15. Cold Spring Harbor Laboratory, Cold Spring Harbor, NY, USA
16. Department of Psychological and Brain Sciences, Dartmouth College, Hanover, NH, USA
17. Institute of Neuroscience, CAS Center for Brain Science, 320 Yue Yang Road Shanghai, 200031 P.R.China; Intelligence Technology, Chinese Academy of Sciences, 319 Yueyang Road, Shanghai 200031, P.R.China
18. Henry H. Wheeler Jr. Brain Imaging Center, Helen Wills Neuroscience Institute, 188 Li Ka Shing Center for Biomedical and Health Sciences, Henry H. Wheeler, Jr. Brain Imaging Center, Suite B107, University of California, Berkeley, CA 94720, USA
19. Center for the Developing Brain, Child Mind Institute, New York, NY; Nathan S. Kline Institute for Psychiatric Research, Orangeburg, NY, USA
20. Simons Collaboration on the Global Brain, Simons Foundation, New York, NY, USA
21. Israel Brain Technologies, Hakfar Hayarok, Ramat Hasharon, Israel
22. Department of Physiology, Keio University School of Medicine, 35 Shinanomachi, Shinjuku-ku, Tokyo,, Japan; RIKEN Brain Science Institute, Laboratory for Marmoset Neural Architecture, 2-1 Hirosawa, Wako, Saitama, Japan
23. Mind Research Network, Department of Electrical and Computer Engineering, University of New Mexico, Albuquerque, NM, USA
24. The Kavli Foundation, Oxnard, CA, USA
25. Johns Hopkins University Applied Physics Laboratory, Laurel, MD, USA
26. Intelligence Advanced Research Projects Activity (IARPA), Maryland Square Research Park, Riverdale Park, MD, USA
27. Center for Imaging Science, Johns Hopkins University, Baltimore, MD, USA
28. Department of Computer Science, Johns Hopkins University, Baltimore, MD, USA
29. Institute for Data Intensive Engineering and Sciences, Johns Hopkins University, Baltimore,MD, USA
30. Department of Neuroscience, Johns Hopkins University, Baltimore, MD, USA
31. Kavli Neuroscience Discovery Institute, Johns Hopkins University, Baltimore, MD, USA
32. Dept of Neuroscience, Baylor College of Medicine, Houston, TX, USA
33. Dept of Molecular Neuroscience, George Mason University, Fairfax, VA, USA
34. Howard Hughes Medical Institute, Rockefeller University, New York, NY, USA
35. Sandia National Laboratories, Albuquerque, NM, USA
36. Harvard Medical School, Harvard University, Boston, MA, USA
37. Department of Molecular and Cellular Neuroscience, The Scripps Research Institute, La Jolla, CA, USA
38. Department of Bioengineering, University of California, San Diego, CA, USA

39. International Brain Research Organization (IBRO), Paris, France
40. Max Planck Institute of Neurobiology, Martinsried, Germany
41. Molecular Biology and Genetic Engineer Institute, CONICET, Argentina
42. Department on Electrical & Computer Engineering, Johns Hopkins University, Baltimore, MD, USA
43. Department of Quantative Neuroscience, University College London, London, England
44. Department of Genetics, Harvard University, Boston, MA, USA
45. U.S. Department of State, Washington D.C., USA
46. Google, Mountain View, CA, USA
47. Department of Statistics, Carnegie Mellon University, Pittsburgh, PA, USA
48. Department of Neurology, Johns Hopkins Hospital, Baltimore, MD, USA
49. National Institute of Neurological Disorders and Stroke (NINDS), National Institute of Health, Bethesda, MD, USA
50. Amazon Web Services, Atlanta, GA, USA
51. Britton Chance Center for Biomedical Photonics, Wuhan National Laboratory for Optoelectronics, Huazhong University of Science & Technology, Wuhan, 430074, China
52. MIT Media Lab, Cambridge, MA, USA
53. Department of Neurology, Johns Hopkins Hospital, Baltimore, MD, USA
54. Office of Naval Research, Arlington, VA, USA
55. Duke Institute for Brain Sciences, Duke University, Durham, NC, USA
56. Department of Neurology, Duke University School of Medicine, Durham, NC, USA
57. National Science Foundation, Arlington, VA, USA
58. Department of Theoretical Physics, MTA Wigner Research Centre for Physics, Budapest, Hungary
59. Department of Psychiatry, Yale University, New Haven, CT, USA
60. Neuroscience Center, University of Geneva, Geneva, Swizterland
61. National Science Foundation, Arlington, VA, USA
62. Salk Institute for Biological Studies, La Jolla, CA, USA
63. Redwood Center for Theoretical Neuroscience, University of California, Berkley, CA, USA
64. Department of Biological Sciences, University of Southern California, Los Angeles, CA, USA
65. Institute for Neuroimaging and Informatics, University of Southern California, Los Angeles, CA, USA